\documentstyle{article}
\begin{document}
%
%
\begin{center}
{\Large \bf
Color Transparency at ELFE\footnote{
Report on the Color Transparency parallel session held
during the second ELFE workshop, Saint Malo, France, September 1996.}}
\vspace{1cm}

Bernard Pire

{\it Centre de Physique Th\'eorique, Ecole Polytechnique, F-91128
 Palaiseau\footnote{Unit\'e propre 14 du CNRS}}

Ingo Sick

{\it Dept of  Physics and Astronomy, Basel University, Basel (CH).
}
\end{center}
\vspace{1cm}

\begin{abstract}
We review the presentations given in the Color Transparency parallel session
 held during the second ELFE workshop.

\end{abstract}

\vspace{1cm}
%
\section{Introduction}

Color Transparency experiments are one of the major goals of the ELFE
project, as
already emphasized\cite{ELFE}.
The concepts behind Color Transparency are now well known: a hard exclusive
scattering
(with a typical large $Q^2$ scale)  selects a very
  special quark configuration in a hadron: the minimal valence state where
all quarks are  close together,   a  small size  color
neutral  configuration  sometimes  referred  to  as a {\em mini hadron}.
Such  a color  singlet system  cannot emit  or absorb  soft gluons
which carry energy or momentum smaller than $Q$.  This is because  gluon
radiation --- like photon radiation in QED --- is a coherent process and
there is thus destructive interference between gluon emission amplitudes
by quarks  with ``opposite''  color. Then, the
 recoiling small components have much reduced strong interactions with
other nucleons
due to this shielding of color.

The first letter of intent\cite{Anklin} emphasized
the measurement of the $(e,e'p)$ cross section on several nuclei and if
possible the
normal component of the recoil polarization. Much work has been  done since
this
time, both theoretically and experimentally\cite{jpr}. Experimental results have
been obtained at SLAC and FNAL, proposals are approved at CEBAF and HERMES,
and new
theoretical ideas have been pursued in many directions. This parallel
session which
 we briefly review has given the opportunity to discuss some of these new
ideas.

\section{Theoretical background}
Strong interaction physics is a fascinating problem and one expects color
transparency
experiments to shed light on some of its aspects. Whether lattice
computations will solve
and explain confinement is still an open question and many
theorists\cite{pen} pursue an effort
to extract from continuum studies some information on basic quantities such
as the gluon
 propagator in a Schwinger-Dyson approach.

How to set the scale of final state interactions in $(e,e'p)$ reactions
needs a careful
examination  of the eikonal approximation in a completely non-relativistic
regime. The Glauber method involves the linearization of the wave equation for
 the ejected proton travelling through the residual nucleus. The
consequences of such
an assumption at high proton momenta has been studied\cite{rad} by comparing
the results with the predictions obtained when the  second-order
differential equation
for the proton wave is solved exactly for each partial wave.

\section{Electroproduction of vector mesons from nuclei}
Diffractive electroproduction at large $Q^2$ involves the pointlike
production of a $q \bar{q}$
pair, its propagation through the medium. Whether the formation of a
genuine vector
meson occurs after or during the travel through the nucleus is of course a
matter of Lorentz
boost. The physical picture of this process has now been developped in many
details and
the presentations of G. Piller\cite{pil} and B. Kopeliovich\cite{kop}
pointed out several
new aspects.

The multichannel evolution equation  approach developed by  B.
Kopeliovich\cite{kop}for the
density matrix describing a hadronic wave packet produced by a virtual
photon and
propagating through  nuclear matter or the vacuum is dual to the quark-gluon
representation.A new procedure of data analysis has been proposed to provide an
unambiguous way of detection of a color transparency signal even at medium
energies.

Using a polarized deuteron offers many advantages: this simplest bound
state of nucleons
has been much studied previously and one may say that its wave function is
well-known up to
$350 MeV$, a limit which may be extended by on-going experiments at NIKHEF,
MAMI and CEBAF.
Moreover, only single and double scattering should be included, the latter
ones probing
smaller $p n$ configurations. Kinematic regions are determined\cite{pil}
where the final state
interaction of the initially produced quark-antiquark pair contributes
dominantly to
 the coherent leptoproduction cross section.

\section{Towards a new experimental proposal}

\renewcommand{\topfraction}{0.8}
\renewcommand{\bottomfraction}{0.8}
\renewcommand{\textfraction}{0.2}
\newcommand{\ct}{colour transparency}
\newcommand{\fsi}{final state interaction}

Much of the literature on \ct\ in the past has concentrated on the aspect
of "transparency", i.e. the reduction of absorption when a nucleon in the
'small' Fock state traverses a nucleus. One of the main difficulties of
this idea relates to the short lifetime of the Fock state selected in a
high-momentum transfer process. The lifetime of this state is typically of
the order of the internucleon distance, not the nuclear size. As a
consequence, much of the \fsi\ of the nucleon asymptotically observed
corresponds to the one of the ordinary nucleon the 'small' Fock state has
evolved back to. The reduction of the overall \fsi\  due to the part of the
trajectory where the nucleon still was in the 'small' state is small,
and correspondingly difficult to disentangle. Increasing the lifetime of the
'small' Fock state by increased momentum (time dilatation) is possible,
but involves enormous momentum transfers, i.e. unpractically small cross
sections.

The contributions of J.M. Laget and D. Voutier presented in this parallel
session\cite{malopar}therefore approach this problem
from a different angle. They chose to study (e,e'p) reactions on the
{\em deuteron}. The emphasis is not on the "transparency", but the
reduction of the \fsi\ between the 'small' Fock state and the other
nucleon\cite{fs}.

In the (e,e'p) process studied, the authors select the kinematics such as to
start from the components of low momentum of the initial deuteron; the wave
function for this configuration is well known. The selection is performed by
ensuring that the scattered electron has an energy that corresponds to the
Bjorken scaling variable $x$=1.  At the same time, the recoiling proton
is selected to correspond to a momentum of the (unobserved) neutron of
$\sim$600MeV/c. This implies that a \fsi\ between the recoiling Fock state
and the neutron with momentum transfer of the order of 600MeV/c had taken
place. The signal for \ct\ is a reduction of the \fsi\ as compared to the one
calculated from the standard N-N interaction.

Laget and Voutier have studied the kinematical range accessible at a
high-energy,
high luminosity facility, and have identified the kinematics where the
standard \fsi\ between proton and neutron has a large effect on the D(e,e'p)
cross section. For this kinematics they then have introduced the change of
the interaction cross section due to \ct, and the effects of the finite
coherence length, using standard parametrizations.

They find, that there are indeed kinematics at large momentum transfers,
Q$^2$=6-16GeV/c$^2$, where the normal nucleonic FSI has a large effect on the
cross section (factors of 2 to 3), and where the \ct\ effect leads to a large
reduction. Given the good control over the deuteron wave function and the
normal, nucleonic \fsi, this change can be expected to be interpretable in an
unambiguous way.

The simulation of potential experiments shows, that for electron energies
in the several - 10GeV range,  for an external high-intensity electron beam
(Luminosity 10$^{38}$ cm$^{-2}$s$^{-1}$) and  spectrometers that can stand
this luminosity, this type of experiment is feasible. Count rates of the order
 of several 10 per hours make this an entirely realistic approach.

The simulations also show that for large-acceptance spectrometers, with their
corresponding much lower luminosity limits, the experiment still is
practical as far as the count rates go. However, the energy resolution does
present problems: to properly measure the missing momentum (i.e. the momentum
transfer in the \fsi) and to guarantee elastic scattering (i.e. absence of
particle production) the resolution of proposed large acceptance
spectrometers such as MEMUS is not good enough; for this option, further work
is required.

\section{Conclusion}
The theory of color transparency is likely to evolve much before ELFE beams
are available
for physics. Experience will be gained from CEBAF and HERMES runs;
understanding in detail
hadron propagation in matter is a challenging goal much related to the
physics of confinement.
%

%
%
\end{document}